\documentclass[doublecol,figures]{epl2} 
\title{Unfolding a composed ensemble of energy spectra
  using singular value decomposition}
\shorttitle{Unfolding an composed ensemble of energy spectra using SVD} 

\author{Richard Berkovits\inst{1}}
\shortauthor{R. Berkovits}% \etal}

\institute{                    
  \inst{1} Department of Physics, Jack and Pearl Resnick Institute, Bar-Ilan University, Ramat-Gan 52900, Israel %\\
}

\abstract{
  In comparing the behavior of an energy spectrum to the predictions
  of random matrix theory one must transform the spectrum such that the
  averaged level spacing is constant, a procedure known as unfolding.
  Once energy spectrums belong to an ensemble where there are large
  realization-to-realization fluctuations the canonical methods for unfolding
  fail. Here we show that singular value decomposition can be used even
  for the challenging situations where the ensemble is composed out of
  realizations originating from a different range of parameters resulting
  in a non-monotonous local density of states. This
  can be useful in experimental situations for which the physical parameters
  can not be tightly controlled, of for situations for which the local
  density of states is strongly fluctuating.
}

\begin{document}

\maketitle

%\section{Introduction}

The statistical behavior of eigenvalues and eigenvectors in random matrix
ensembles has a rich and distinguished history in enhancing our understanding
of various quantum systems, ranging from nuclear spectra to quantum gravity
\cite{wigner51,dyson62,porter65,gorkov65,bohigas84,david85,altshuler86}.
The random matrix
ensemble  \cite{ghur98,alhassid00,mirlin00,evers08,mehta91}
is defined by the probability of a matrix $M$ being given by
$P_a(M)=\exp(-a~\mathrm{Tr}~M^2)$, where $a$ is a positive constant
determining the distribution's width. This is equivalent to independently
drawing the matrix elements $M_{i,j}$ from a distribution with zero
average and variance determined by $a$.

The eigenvalues $E_i$ and eigenvectors $\psi^i_j$ follow random matrix theory
(RMT). RMT predictions are based on the assumption that the average level
spacing $\Delta_i = \langle E_i - E_{i-1} \rangle$ is constant, where averaging
is over all realizations belonging to the ensemble.
However, since the level density of the matrix
follows a semi-circle law, i.e., is not constant, the eigenvalue spectrum must
be unfolded using the averaged level density to obtain a constant level
spacing.
For any well-behaved distribution with finite variance, an ensemble
average of the local density of states can be performed to extract the
local level spacing for unfolding. After unfolding, the constant $a$ has
no effect on the spectra's properties and RMT predictions become universal.

Recently, attention has been drawn to ensembles that are expected physically
to exhibit pure chaotic behavior (i.e., RMT statistics) while the corresponding
matrix is sparse \cite{bohigas71}.
One such model is the Sachdev-Ye-Kitaev model
\cite{sachdev93,kitaev15,sachdev15,maldacena16}. Due to the fewer
independent random elements in the matrix compared to a canonical random
matrix, there are strong fluctuations from one realization to another.
This renders the system non-self-averaging, and thus, the standard unfolding
method will not work, and a more careful unfolding method
must be applied \cite{sonner17,gharibyan18,jia20,berkovits23}.

Combining realizations from different RMT ensembles into a single
ensemble can give rise to similar behavior.
Consider a situation where eigenvalues for multiple systems are obtained, each
drawn from different ensembles but still following RMT predictions. If the
origin of each spectrum can be determined, each can be unfolded according
to its corresponding ensemble, revealing that all the spectra follow RMT
predictions.
However, identifying the origin of a particular spectrum may
not be straightforward. In this context, singular value decomposition offers
a straightforward and efficient solution to unfolding spectra from different
ensembles without prior labeling.
This could have significant relevance in the analysis of experimental
spectra from systems where the conditions are poorly controlled or in
numerical studies aiming to expand the ensemble by combining results
obtained from calculations performed under different conditions, such
as varying system size.

In this letter, we will examine several example of a composed ensemble.
The first is an extension of the canonical RMT ensemble, where the value
of $a$ is drawn from a distribution $\tilde P(a)$ with finite variance,
and $M$ is then drawn from the conditional probability
$P(M|a)=\exp(-a~\mathrm{Tr}~M^2)$. The combined probability is given
by $P(a,M)=\tilde P(a) P_a(M)$, and $P(M)$ is obtained by integrating $P(a,M)$
over $a$. If a specific value of $a_0$ is chosen
(i.e., $\tilde P(a)=\delta(a-a_0)$), the ensemble will follow RMT.
This should hold for any value of $a$ in the distribution, as long as the
eigenvalues of $M$ are unfolded according to the value of $a$ from which $M$
was drawn. Naturally, this composed ensemble is governed by a single parameter.
A more heterogeneous composed ensemble is built by choosing realizations
of different matrix sizes in addition to different $a$. Finally an ensemble
is composed out of eigenvalues belonging to different regions of the band
as well as from different realizations corresponding to the distribution
of variances $P(a)$.
Thus, although each sequence of eigenvalues follows RMT behavior, unfolding
becomes quite challenging.

The fact that all these composed ensembles follow RMT can be most clearly
seen by using measures that avoid unfolding.
Such a measure for
short energy scale statistics is the
ratio statistics \cite{oganesyan07}, defined as:
\begin{eqnarray} \label{ratio}
r_s &=& \langle \min (r_i,r_i^{-1}) \rangle_i,
\\ \nonumber
r_i &=& \frac {E_i-E_{i-1}}{E_{i+1}-E_{i}},
\end{eqnarray}
where the average is over the ensemble and levels.
For the Wigner  distribution $r_s \cong  0.5307$
for the GOE symmetry \cite{atas13}. Since this measure is based on ratios
between consecutive level spacings there is no need to unfold. Thus, it
is ideally suited to verify that RMT predictions are followed, at least for
short energy scales.
Indeed, for the first composed ensemble,
$r_s$ is numerically calculated
for $4000$ real symmetric
matrices of sizes $16000 \times 16000$,
and $32000 \times 32000$. For each realization a random value
$W$  is drawn from a box distribution between
$2-Y \ldots 2+Y$, then
the off-diagonal elements $M_{i,j}$ are
drawn from a box distribution between $-W \ldots W$.
The matrices are exactly diagonalized and the eigenvalues are obtained.
Thus, $W$, determines the variance of the off-diagonal
elements. 
Three values of $Y=0,1,2$ are considered. For $Y=0$ one obtains
the the usual RMT ensemble, while for $Y=1$ and $Y=2$ one gets a
composed ensemble with
a different variance of the matrix elements depending on the $W$ drawn, i.e.,
an composed ensemble.
The ratio statistics for different sizes, and different values of $Y$ are 
calculated resulting for all the cases in $r_s =  0.5307 \pm 0.0002$.
Thus, neither size nor 
$Y$, have influence on $r_s$. Therefore, the value of $r_s$ remains unaffected
by both the size and $Y$, indicating that the composed ensemble conforms to
the predictions of RMT for the ratio statistics. These findings are consistent
with the behavior of other composed ensembles discussed in the
latter part of the letter.

%\begin{figure}
%\onefigure[width=8cm]{ergxrs.eps}
%\caption{The ratio statistics, $r_s$, for different sizes of matrices
%  ($8000 \times 8000$, $16000 \times 16000$, and $32000 \times 32000$)
%  and different composed ensembles ($Y=0,1,2$). At infinite size of matrices
%one expects $r_s \cong  0.5307$.}
%\label{fig1}
%\end{figure}

For the study of larger energy scales unfolding is crucial. One of the
earliest measures used to probe the long range behavior of the
spectrum is the variance of the number of levels as function
of the size  of an energy window $E$, expressed by
$\langle \delta^2 n(E) \rangle$ (here $n(E)$ is the number of levels in the
window). RMT predicts that the variance grows logarthimically
(for GOE $\langle \delta^2 n(E) \rangle =
(2/\pi^2)\ln(\langle n(E) \rangle)+0.44$).
A departure from the RMT behavior will manifest in a stronger
than logarithmic increase of the variance.
The energy of departure is identified with the Thouless energy and is not
expected for a pure RMT ensemble.
Thus, we expect that the number variance will not depart from the
logarithmic behavior at any energy.

\begin{figure}
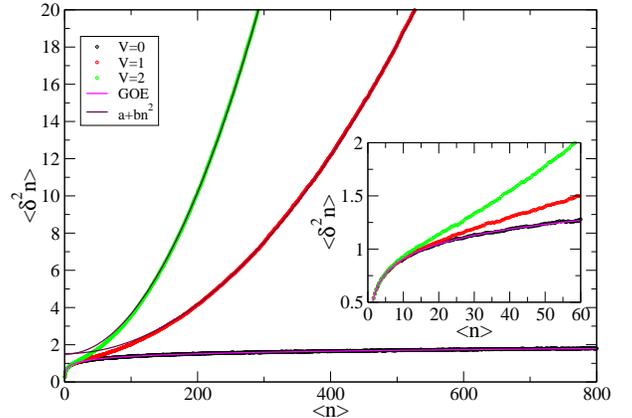

  \onefigure[width=8cm]{ergxvn.eps}
\caption{  The level number variance, $\langle \delta^2 n(E) \rangle$, as
  function of the energy window $E$. Symbols represent
  the numerical results for the different composed ensembles
  ($Y=0$, black ; $Y=1$, red and $Y=2$, green) with local ensemble unfolding.
The GOE prediction ($\langle \delta^2 n(E) \rangle =
(2/\pi^2)\ln(\langle n(E) \rangle)+0.44$) corresponds to the magenta line.
Fits for larger values of $\langle n \rangle$ by $a+b\langle n(E) \rangle^2$
for $Y=1,2$ are represented by curves.
Insert: zoom into smaller values of $\langle n \rangle$.
Deviation from the GOE behavior for $Y=1,2$ 
  are observed even for small values of $\langle n \rangle$.
}
\label{fig2}
\end{figure}

For the number variance, one can not avoid unfolding. The simplest form
of unfolding is given by the local ensemble unfolding, where
the nearest neighbour level spacing is averaged over the realizations
belonging to the (composed) ensemble in a given energy region and then
it is possible to reconstruct a given realization
spectrum such that the level spacing is on the average equal to one everywhere.
Specifically, the i-th level averaged spacing
is $\Delta_i = \langle E_{i+p}-E_{i-p}\rangle/2p$
(the results do not strongly depend on $p$. For all further calculations
$p=5$ was chosen), and the unfolded spectrum for the j-th realization is
$\varepsilon^j_i=\varepsilon^j_{i-1}+(E^j_i-E^j_{i-1})/\Delta_i$.
In order to improve statistics we average  also
over the center of the energy window. Here
the number variance is also averaged over $21$ positions of the
center of the energy window, $\tilde E$,
equally spaced around the band center, where the
furthest point is no more than $1/15$ of the bandwidth from the center.
The number of states, $n(E,\tilde E)$, in a window of width $E$ centered
at $\tilde E$, 
is calculated, then the averages $\langle n(E) \rangle$
and $\langle n^2(E) \rangle$ are taken over all positions of the center
$\tilde  E$ and all realizations.
Results of calculating the level number variance of the unfolded
eigenvalue spectrum are presented in Fig. \ref{fig2}.

As expected the canonical ensemble ($Y=0$) fits very well the RMT
predictions, even for large energy scales. On the other hand, for a
composed ensemble the variance diverges from the RMT predictions at
rather small values of $\langle n \rangle$. At larger values it fits quite
well a quadratic form $a + b \langle n \rangle$
(with $b= 6.6 \times 10^{-5}$ for $Y=1$ and $b=2.2 \times 10^{-4}$ for $Y=2$).
The quadratic behavior could be understood assuming that the averaged
level spacing for a matrix with matrix element variance $W$ is
$\Delta_W=\Delta+c \sqrt{W-2}$, where $\Delta$ is the
level spacing averaged over the whole composed ensemble
and $c$ is a constant. 
Thus,
$\langle \delta^2 n \rangle =
\int dW \tilde P(W) \langle (n_W -\langle n \rangle)^2 \rangle$,
taking into account
that after unfolding $\langle n \rangle = n$ and
$n_W \sim n (1+c \sqrt{W-2}/\Delta)$,
results in $\langle \delta^2 n \rangle \sim
\int dW \tilde P(W) (W-2) (c n/ \Delta)^2$, leading
for the box distribution to
$\langle \delta^2 n \rangle \sim
\int_{2-Y}^{Y+2} dW (W-2) (c n/ \Delta)^2$.
Thus, $b= Y^2 (c / \sqrt{2} \Delta)^2$, i.e., $b \propto Y^2$,
which is in line with the $b$ values quoted above.

Thus, when analyzing long-range energy spectra properties in
composed ensembles ($Y>0$), it has been observed that using
local ensemble unfolding results in divergence from RMT predictions.
In order to accurately identify and analyze these properties,
a different unfolding technique is necessary. This is where the
singular value decomposition (SVD) comes in. Multiple studies
\cite{fossion13,torres17,torres18,berkovits20,berkovits21,berkovits22,rao22,rao23,berkovits23}
have demonstrated the power of SVD unfolding.
%can be used to characterize the features of the spectrum, for example
%on what scale does it follow RMT predictions
%
For the SVD analysis, $P$ eigenvalues around the center of the band
of $M$ realizations of disorder are written down 
as a matrix $X$ of size $M \times P$ where $X_{mp}$ is the $p$-th level of
the $m$-th realization. The matrix
$X$ is decomposed to a multiplication of three matrix $X=U \Sigma V^T$, where
$U$ is a $M\times M$ matrix and $V$ is a $P \times P$ matrix
while $\Sigma$ is a $M \times P$ diagonal matrix of rank
$r=\min(M,P)$. The $r$ diagonal elements of $\Sigma$
are the singular values amplitudes $\sigma_k$ of the matrix $X$.
$\sigma_k$ are always positive
and may be ordered from the largest to the smallest
$\sigma_1 \geq \sigma_2 \geq \ldots \sigma_r$, the square
of the singular values are defined as $\lambda_k=\sigma_k^2$.
The Hilbert-Schmidt (Frobenius) norm of the matrix
$||X||_{HS}=\sqrt{Tr X^{\dag}X}=\sum_k \lambda_k$.
Using $U$ and $V$ one can define an auxiliary matrix $X^{(k)}_{ij}=U_{ik}V^T_{jk}$
which may be used to express $X$ as a sum of these auxiliary matrices, resulting
in $X_{ij}=\sum_k \sigma_k X^{(k)}_{ij}$. Truncating the sum after $m$ terms
(modes) will result in an approximation to the matrix
$\tilde X =\sum_{k=1}^m \sigma_k X^{(k)}$, for which $||X||_{HS}-||\tilde X||_{HS}$
is minimal. To write down the matrix containing $P$ eigenvalues for the $M$
realizations one requires $MP$ independent
variables. Using the approximate matrix $\tilde X$, $m(M+P)$ independent
variables are needed, i.e., for small $m$ much less information.

The essence of unfolding is removing global features of the energy spectrum,
so the averaged local spacing is constant (unity). For the
local ensemble average one unfolds using a simple average over
the whole ensemble, thus employing $P$ independent values to describe the
average spectrum which is used to unfold. SVD is more subtle. For
simplicity lets consider approximating the spectrum by only the first mode
of the SVD ($m=1$). Explicitly, $X^{(1)}_{ij}=U_{i1}V^T_{j1}$ and
$\tilde X_{ij} = \sigma_1 X^{(1)}_{ij}$. Thus, the $P$ values of $V^T_{j1}$
common to all realizations, essentially capturing an average spectrum
for all $M$ realizations, are multiplied for each realization $i$ with
a scale factor  $U_{i1}$, thus requiring $M+P$ parameters for the description
of the global properties of the energy spectrum.

\begin{figure}
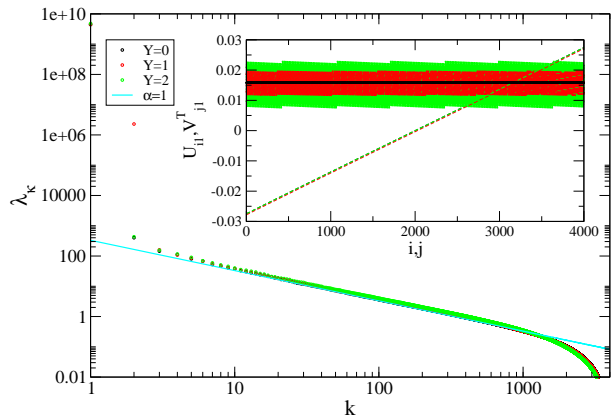

\onefigure[width=8cm]{ergxsvd.eps}
\caption{A scree plot of
  the squared singular values, $\lambda_k$ as function of the
  mode number $k$ for a $16000 \times 16000$  matrix where $P=4000$ eigenvalues
  at the middle of the band for $M=4000$ realizations are considered. As
  expected for Wigner statistics $\lambda_k \sim k^{-1}$
  for the regular ($Y=0$) and composed ($Y=1,2$) ensembles,
  except for $k=1,2$.  
  Inset: $V^T_{j1}$, which is proportional to the averaged eigenvalues over
  the composed ensemble, and the scale factor for the i-th
  realization  $U_{i1}$. While for $Y=0$, $U_{i1}$ is almost constant, for
  $Y=1,2$, shows sample to sample fluctuations. 
}
\label{fig3}
\end{figure}

To illustrate the SVD unfolding of a composed ensemble we start
by calculating
the square of the singular values, $\lambda_k$, for a regular ($Y=0$) and
a couple of composed ($Y=1,2$) ensembles. The results are presented in Fig.
\ref{fig3} where the squared singular values, $\lambda_k$ as function of the
mode number $k$ is shown for the (composed) ensembles of $M=4000$
different disorder realizations of
$16000 \times 16000$  matrices. $P=4000$ eigenvalues
at the middle of the band for $M=4000$ realizations are written into a
$4000 \times 4000$ matrix $X$ and SVD is performed. As
expected for Wigner statistics the singular values squared follow
a power law $\lambda_k \sim k^{-1}$ \cite{fossion13,torres17,torres18}.
for all regular ($Y=0$) and composed ($Y=1,2$) ensembles except for
the lowest modes, $k=1,2$. One would assume that the averaged eigenvalue
spectrum will be similar for any value of $Y$. Indeed, examining the
lowest mode ($k=1$) can provide us with some insight into
how SVD performs the unfolding of the spectrum.
The j-th averaged eigenvalues over the whole composed ensemble,
$V^T_{j1}$, is almost independent of $Y$, which indeed confirms our
expectations
(see the inset of Fig. \ref{fig3}).
The distinction between various values of $Y$ becomes apparent when analyzing
the scale factor ($U_{i1}$) for each realization, as shown in the inset
of Fig. \ref{fig3}.
While for $Y=0$, $U_{i1}$ almost does not depend on $i$, for
$Y=1,2$, strong sample to sample fluctuations in $U_{i1}$ are obvious.
Thus, SVD essentially adjusts the averaged eigenvalues to each particular
realization by multiplying it with a scale factor.

\begin{figure}
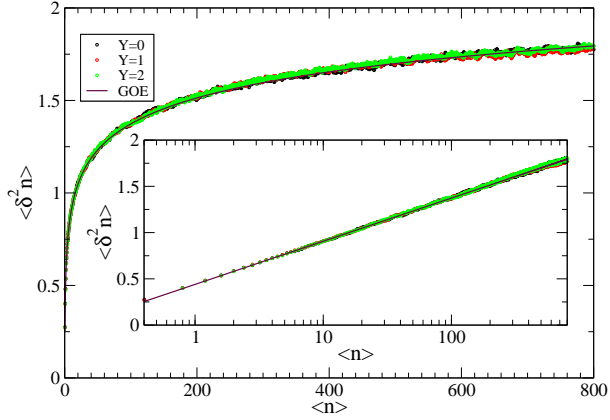

\onefigure[width=8cm]{ergxvecvn.eps}
\caption{
  The level number variance, $\langle \delta^2 n(E) \rangle$, as
  function of the energy window $E$. Symbols represent
  the numerical results for the different composed ensembles
  ($Y=0$, black ; $Y=1$, red and $Y=2$, green) for SVD ensemble unfolding.
The GOE prediction ($\langle \delta^2 n(E) \rangle =
(2/\pi^2)\ln(\langle n(E) \rangle)+0.44$) is depicted by the magenta line.
Insert: A semi-log plot of the same data, which shows that
the regular ($Y=0$) and composed ($Y=1,2$) ensembles fit GOE perfectly.
}
\label{fig4}
\end{figure}

The SVD unfolding technique employs a level spacing that is specific to
each realization. In particular, the level spacing of the j-th realization
for the i-th energy level is given by:
$\Delta^j_i = (\tilde \varepsilon^j_{i+p}-\tilde \varepsilon^j_{i-p})/2p$, where
$\tilde \varepsilon^j_i=\sum_{k=1}^2 \sigma_k U_{ik}V^T_{jk}$.
The number variance after performing the SVD unfolding is presented
in Fig. \ref{fig4}. The number variance, $\langle \delta^2 n(E) \rangle$,
fits perfectly the GOE predictions and is independent of $Y$.
Thus, realization-to-realization
fluctuations endemic to the composed ensembles are
completely eliminated by the SVD unfolding. 

\begin{figure}
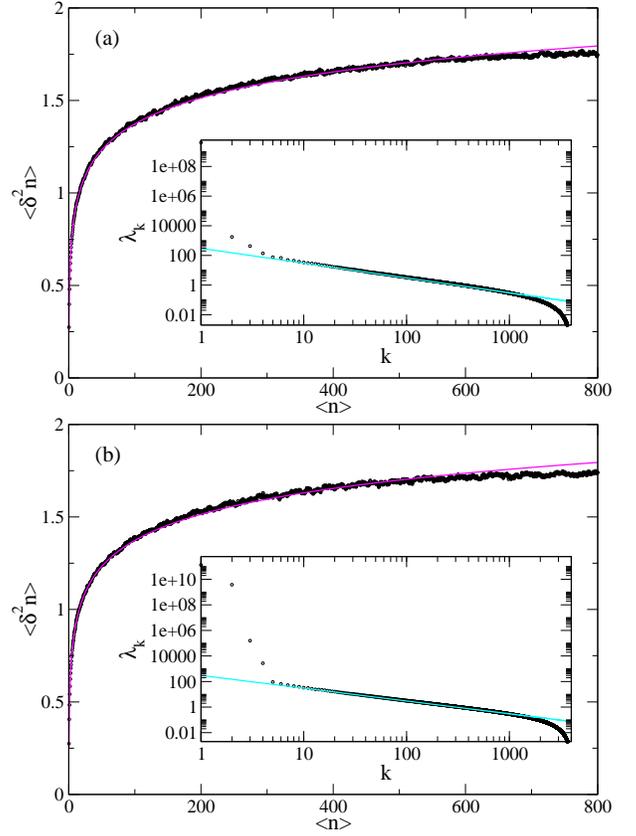

\onefigure[width=8cm]{ergxll.eps}
\onefigure[width=8cm]{ergxlx.eps}
\caption{
  The level number variance, $\langle \delta^2 n(E) \rangle$, as
  function of the energy window $E$ for $P=4000$ eigenvalues and
  $M=4000$ realizations. Symbols represent
  the numerical results for (a) $Y=2$ and two different sizes
  $16000 \times 16000$ and $L=32000 \times 32000$ where the
  eigenvalues are chosen from the middle of the band, (b)
  $Y=2$ with a single size $32000 \times 32000$ where a block of
  $4000$ consecutive
  eigenvalues are chosen for each realization at random from different regions
  of the band. The GOE prediction are depicted by the magenta line.
  Inset: A scree plot of
  the squared singular values, $\lambda_k$ as function of the
  mode number $k$. The cyan line corresponds to a $k^{-1}$ power law.
}
\label{fig5}
\end{figure}

Let's now see what happens as we challenge SVD unfolding with a more
demanding composed ensemble. For the previous example of a composed ensemble
one may argue that it is not surprising that the SVD unfolding works so well
for an ensemble for which the variance of the matrix term is not identical for
different realizations, since the spectrum is governed by a single parameter,
essentially the width of the semi-circle eigenvalue density. Can it work for
more complicated situations? Here we examine two such composite ensembles.

The first is constructed by adding realizations which are drawn from 
different variance of the matrix elements (specifically $Y=2$) and different
matrix sizes (here $16000 \times 16000$ and $32000 \times 32000$).
Thus, for each realization added to the composite ensemble we first
draw a size (equal probability for the two sizes) and then $W$ in the
range $2-Y \ldots 2+Y$. Since we take $P=4000$ consecutive eigenvalues
straddling the center, these eigenvalues for the different sizes cover
different regions of the band and therefore cannot be captured by a single
parameter. The SVD amplitudes for a composite ensemble of $M=4000$ realizations 
are presented in the inset of Fig. \ref{fig5}a. For $k>4$, the amplitudes follow
a power law $\lambda_k \sim k^{-1}$ as expected from GOE. Compared to the
fixed size composite ensemble (Fig. \ref{fig3}), it is not surprising that more
modes are needed to capture the sample to sample fluctuations. Thus, here
the unfolding features
$\tilde \varepsilon^j_i=\sum_{k=1}^4 \sigma_k U_{ik}V^T_{jk}$, leading
to the number variance presented in Fig. \ref{fig5}a. GOE behavior fits
quite well up to $\langle n(E) \rangle \sim 600$, where
$\langle \delta^2 n(E) \rangle$ goes below the GOE prediction. A similar
behavior has been seen for the Sachdev-Ye-Kitaev model with
SVD unfolding \cite{berkovits23} and other unfolding methods \cite{jia20}.

The second composed ensemble chooses realizations with a different variance of
the matrix elements (again $Y=2$), though of the same size
($32000 \times 32000$). The  center of the $P=4000$
consecutive eigenvalues is chosen at random for each realization in
the range $4000 \ldots 28000$. Thus, different band regions for each realizations
are sampled, and the since the eigenvalues are not at the center of the band
their density is skewed. Nevertheless, the SVD unfolding seems to do a decent
work, as can be seen in Fig. \ref{fig5}b. Again, the SVD amplitudes 
exhibit a power law $\lambda_k \sim k^{-1}$ for $k>4$. Unlike for the previous
composed ensemble where the $k=1$ amplitude was orders of magnitude above
all the other modes, here both $k=1$ and $k=2$ are orders of magnitude above
the rest (inset Fig. \ref{fig5}b), indicating that the unfolding requires
more than a simple single
parameter scaling. Nevertheless, the SVD unfolding results in a fit to the GOE
as long as  $\langle n(E) \rangle < 500$.

In conclusion, SVD unfolding allows for the recovery of RMT statistics on
large energy scales for an ensemble composed of
realizations drawn from different RMT ensembles, which
correspond to varying level density profiles of their spectrum. Despite
the absence of tags indicating the origin of the spectrum and strong
realization-to-realization fluctuations, SVD accurately identifies the
appropriate level spacing for unfolding. It would be intriguing to
further investigate the limitations and applicability of SVD unfolding
in more complex composed ensembles with diverse underlying symmetries
(Poisson, GOE, GUE, GSE). Additionally, exploring alternative spectral
unfolding techniques, such as machine learning, could provide valuable
insight into their effectiveness and efficiency when compared to SVD unfolding.

%------------------------------------------>>>>

%% here a revision

%\revision{Insert here the text.
%See fig.~\ref{fig.1}, table~\ref{tab.1} and eq.~(\ref{eq.1}).
%See also~\cite{b.a,b.b}.}

%here a shortcut $\emc$ and again $\emc$

%\begin{equation}
%\label{eq.1}
%0\neq1
%\end{equation}

%\begin{table}
%\caption{Table caption.}
%\label{tab.1}
%\begin{center}
%\begin{tabular}{lcr}
%first  & table & row\\
%second & table & row
%\end{tabular}
%\end{center}
%\end{table}

%\acknowledgments
%Insert here the text.


\begin{thebibliography}{0}

\bibitem{wigner51}
  \Name{Wigner, E.P.}
  \REVIEW{Ann. Math.}{53}{1951}{36}.
  \SAME{62}{1955}{548}.
  \SAME{65}{1957}{203}.
  \SAME{67}{1958}{325}.

\bibitem{dyson62}
  \Name{Dyson, F.J.}
  \REVIEW{J. Math. Phys.}{3}{1962}{140}.
  \SAME{3}{1962}{157}.

\bibitem{porter65}
  \Name{Porter, C.E.}
  \Book{Statistical Theory of Spectra: Fluctuations}
%  \Editor{A. Editor}
%  \Vol{9}
  \Publ{Academic, New York}
  \Year{1965}
%  \Page{666}.

\bibitem{gorkov65}
  \Name{Gor’kov, L. P., \and Eliashberg G. M.}
  \REVIEW{Sov. Phys. JETP}{21}{1965}{940}.

\bibitem{bohigas84}
  \Name{Bohigas, O., Giannoni M.-J. \and Schmit C.}
  \REVIEW{Phys. Rev. Lett.}{52}{1984}{1}.

\bibitem{david85}
  \Name{David F.}
  \REVIEW{Nucl. Phys. B}{257}{1985}{543}.
  
\bibitem{altshuler86}
  \Name{Altshuler B. \and Shklovskii B.}
  \REVIEW{Sov. Phys. JETP}{64}{1986}{127}.


%F.J. Dyson, M.L. Mehta, J. Math. Phys. 4 (1963) 701.

\bibitem{ghur98}
  \Name{Guhr T.,Muller-Groeling A. \and Weidenmuller H. A.}
  \REVIEW{Phys. Rep.}{299}{1998}{190}.

\bibitem{alhassid00}
  \Name{Alhassid Y.}
  \REVIEW{Rev. Mod. Phys.}{72}{2000}{895}.

\bibitem{mirlin00}
  \Name{Mirlin A.D.}
  \REVIEW{Phys. Rep.}{326}{2000}{259}.

\bibitem{evers08}
  \Name{Evers R. \and Mirlin A.D.}
  \REVIEW{Rev.  Mod. Phys.}{80}{2008}{1355}.

\bibitem{mehta91}
  \Name{Mehta M. L.}
  \Book{Random matrices}
  \Publ{Acad. Press, New York}
  \Year{1991}.

\bibitem{bohigas71}
  \Name{Bohigas O. \and Flores J.}
  \REVIEW{Phys. Lett. B}{34}{1971}{261}.

\bibitem{sachdev93}
  \Name{Sachdev S. \and Ye J.}
  \REVIEW{Phys. Rev. Lett.}{70}{1993}{3339}.
  
\bibitem{kitaev15}
  \Name{Kitaev A.}
  \Book{Talks at the KITP on April 7th and May 27th,
    http://online.kitp.ucsb.edu/online/entangled15/kitaev/,
    http://online.kitp.ucsb.edu/online/entangled15/kitaev2/}
  \Publ{KITP, Santa Barbara}
  \Year{2015}.

\bibitem{sachdev15}  
  \Name{Sachdev S.}
  \REVIEW{Phys. Rev. X}{5}{2015}{041025}.

\bibitem{maldacena16}
  \Name{Maldacena J. \and Stanford D.}
  \REVIEW{Phys. Rev. D}{94}{2016}{106002}.

%\bibitem{bohigas84}
%  \Name{Bohigas, O., \and M.-J. Giannoni}
%  \Book{Mathematical and Computational Methods in Nuclear Physics}
%  \Editor{J.S. Dehesa, J.M.G. Gomez, and A. Polls}
%  \Vol{209}
%  \Publ{Publishing house, City}
%  \Year{1984}
%  \Page{1}.

\bibitem{sonner17} 
  \Name {Sonner J. \and M. Vielma M.}  
  \REVIEW{J. High Energ. Phys.}{11}{2017}{149}.  


\bibitem{gharibyan18}
  H. Gharibyan, M. Hanada, S. H. Shenker and M, Tezuka,
  J. High Energ. Phys. {\bf 07}, 124  (2018).

\bibitem{jia20}
  \Name {Jia Y. \and Verbaarschot J.J.M.}
  \REVIEW{J. High Energ. Phys.}{07}{2020}{193}.  

\bibitem{berkovits23}
  \Name{Berkovits R.}
  \REVIEW{Phys. Rev. B}{107}{2023}{035141}.

\bibitem{oganesyan07} 
  \Name{Oganesyan V. \and Huse D. A.}
  \REVIEW{Phys. Rev. B}{75}{2007}{155111}.

\bibitem{atas13} 
  \Name{Atas Y. Y., Bogomolny, E., Giraud O. \and Roux G.}
  \REVIEW{Phys. Rev. Lett.}{110}{2013}{084101}.

\bibitem{fossion13}
  \Name{Fossion R., Torres-Vargas G. \and L\'opez-Vieyra J. C.}
  \REVIEW{Phys. Rev. E} {88}{2013}{060902(R)}.

\bibitem{torres17}
  \Name{Torres-Vargas G., Fossion R., Tapia-Ignacio C. \and L\'opez-Vieyra J. C.}
  \REVIEW{Phys. Rev. E}{96}{2017}{012110}.

\bibitem{torres18}
  \Name{Torres-Vargas G., M\'endez-Berm\'udez J. A., L\'opez-Vieyra J. C. \and
    Fossion R.}
  \REVIEW{Phys. Rev. E}{98}{2018}{022110}.

\bibitem{berkovits20}
  \Name{Berkovits R.}
  \REVIEW{Phys. Rev. B}{102}{2020}{165140}.

\bibitem{berkovits21}
  \Name{Berkovits R.}
  \REVIEW{Phys. Rev. B}{104}{2021}{054207}.

\bibitem{berkovits22}
  \Name{Berkovits R.}
  \REVIEW{Phys. Rev. B}{105}{2022}{104203}.

\bibitem{rao22}
  \Name{Rao W.-J.}
  \REVIEW{Phys. Rev. B}{105}{2022}{054207}.

\bibitem{rao23}
  \Name{Xu W.-F. \and Rao W.-J.}
  \REVIEW{Sci. Rep.}{13}{2023}{634}.

\end{thebibliography}
\end{document}